\documentclass[twocolumn,preprintnumbers,amsmath,amssymb]{revtex4}
\usepackage{graphicx}% Include figure files
\usepackage{dcolumn}% Align table columns on decimal point
\usepackage{bm, amsmath, amssymb}% bold math
\newcommand{\comment}[1]{}

\newcommand{\teff}[0]{T_\textrm{eff}}
\newcommand{\vc}[0]{V_\textrm{c}}
\newcommand{\dvc}[0]{\Delta V_\textrm{c}}

\begin{document}

\title{Quantum Fluctuations in the Chirped Pendulum}% Force line breaks with \\

\author{K. W. Murch$^1$, R. Vijay$^1$,  I. Barth$^2$, O. Naaman$^1$, J. Aumentado$^3$, L. Friedland$^2$, I. Siddiqi$^1$}

\date{\today}% It is always \today, today,
             % but any date may be explicitly specified

%\begin{abstract}
\address{$^1$ Quantum Nanoelectronics Laboratory, Department of Physics, University of California, Berkeley, California 94702, USA\\ $^2$Racah Institute of Physics, Hebrew University, Jerusalem 91904, Israel\\
$^3$National Institute of Standards and Technology, 325 Broadway, Boulder, Colorado 80305, USA}

\maketitle

\textbf{Anharmonic oscillators, such as the pendulum, are widely used for precision measurement \cite{bakependulum} and to model nonlinear phenomena \cite{khal96book}. In any physical detector, fluctuations\textemdash such as thermal or quantum mechanical noise\textemdash can excite random motion in the oscillator, ultimately imposing a bound on measurement sensitivity. In systems where equilibrium is established with the noisy environment, for example most electrical or optical amplifiers, noise-induced broadening scales with the intensity of fluctuations, which for thermal noise is proportional to temperature ($T$). But how does noise affect an out of equilibrium oscillator where the motion is varied faster than energy is exchanged with the environment? We create such a scenario by applying fast, frequency chirped voltage pulses to a nonlinear superconducting resonator where the ring down time is longer than the pulse duration.  Under these conditions, the circuit oscillates with either small or large amplitude depending on whether the drive voltage is below or above a critical value--a phenomenon known as autoresonance \cite{faja01ajp}. This sharp threshold is significant in planetary dynamics \cite{malh93pluto} and plasmas \cite{faja99plasma}, enables the excitation of particles in cyclotron accelerators \cite{livi54book}, and may even be used to detect the state of a quantum two level system \cite{naam08chirp}. For the first time, we observe a noise-induced broadening of the autoresonant threshold that is $\propto \teff^{1/2}$ where $\teff = T$ in the classical regime and $\hbar\omega/2k_B$ in the quantum regime. The results imply that fluctuations only determine the initial conditions of such a non-equilibrium oscillator and do not effect its time evolution.}

Our oscillator is a superconducting electrical circuit based on a Josephson tunnel junction formed by two thin aluminum films separated by an oxide barrier. In response to a periodic current drive $I(t)$, the junction voltage $V(t)$ oscillates according to the Josephson relations, which describe the motion of a particle with coordinate $\delta$ in a sinusoidal potential $U(\delta) =  [ \hbar I_0/(2e)] \cos(\delta)$, where $I_0$ is the junction critical current, $\delta$ is the phase difference across the junction, and $e$ is the electron charge.  This system is analogous to a mechanical pendulum with angular coordinate $\theta = \delta$
(Fig.\ 1a and b). The response of such an anharmonic oscillator is illustrated in Figure 1c; as the drive strength is increased the resonant response shifts to lower frequency and ultimately exhibits bistability. In this regime, if the oscillator is excited with a constant frequency drive of increasing amplitude, as depicted by the light grey arrow in Figure 1c, it abruptly switches from a low to a high amplitude oscillation state. This effect has been used to realize a threshold amplifier \cite{sidd04rfjba} to readout the state of quantum bits \cite{sidd06jba,lupa06,mall09}.  In the presence of fluctuations, the switching from one oscillation state to the other is modeled as an equilibrium Arrhenius process involving escape over an effective energy barrier \cite{dykm88,dykm07}. Consequently, the sensitivity of this amplifier improves with decreasing temperature until the quantum regime is reached \cite{vija09rev}.

    \begin{figure*}
\begin{center}
\includegraphics[angle = 0, width = .9\textwidth]{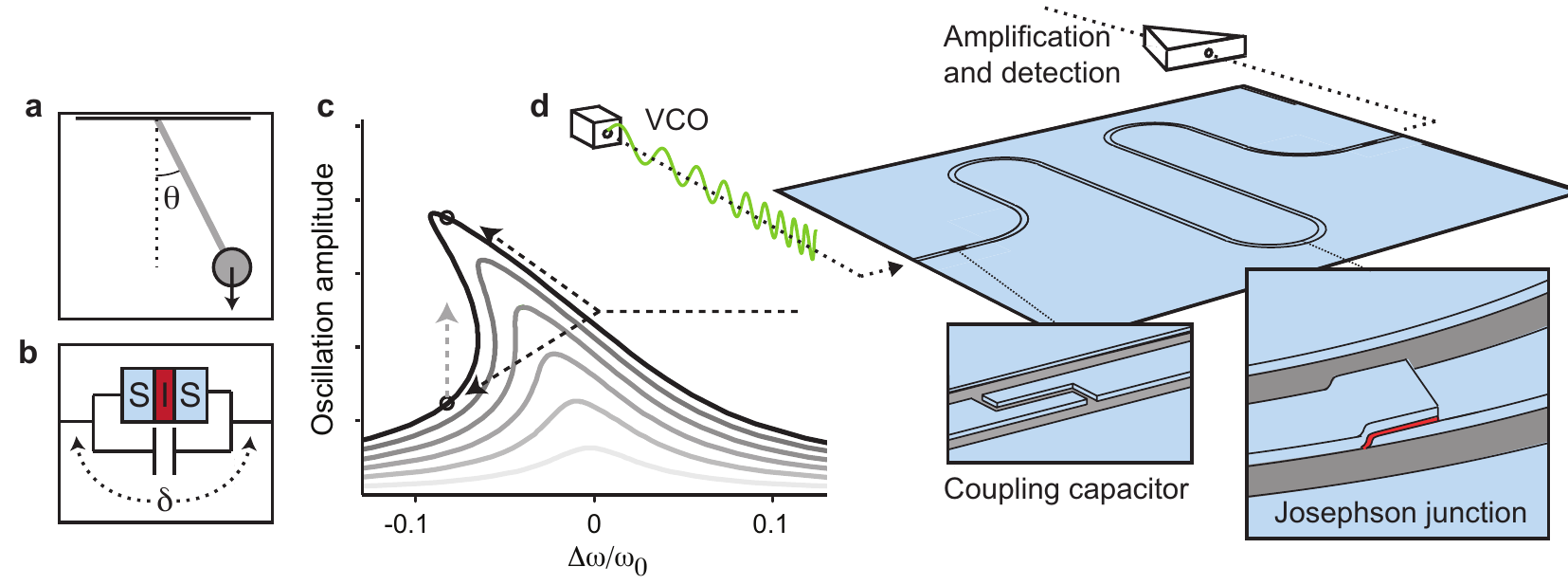}
\end{center}
\caption[setup]{{\bf The pendulum.} The resonant response of an anharmonic oscillator, as realized by a pendulum ({\bf a}) or a capacitively shunted Josephson tunnel junction ({\bf b}), becomes increasingly asymmetric ({\bf c}) as the drive strength is increased.  Above a critical excitation the oscillator exhibits two metastable states, as indicated by black circles, with low and high amplitude of oscillation.  These states may be excited by ramping the drive amplitude at a fixed frequency (grey dashed arrow) or by chirping the drive at fixed power (black dashed arrow).  {\bf d}, Schematic of the oscillator circuit consisting of a superconducting resonator embedded with a Josephson junction. The frequency modulated drive signal is generated using a voltage controlled oscillator (VCO) and detected by means of a cryogenic amplifier and room temperature electronics (not shown).}
\end{figure*}

The bistable regime may also be accessed by driving the anharmonic oscillator with a fixed amplitude, chirped drive as depicted by the black arrows in Figure 1c. For a given chirp rate $\alpha$ and oscillator quality factor $Q$, if the drive amplitude is above a critical voltage $V_{C} \propto \alpha^{3/4}$, the oscillator phase locks with the drive signal and climbs to the high amplitude oscillation state \cite{faja01ajp}. Although known in plasma and accelerator physics since the 1950s, autoresonance has only recently been observed in an electrical circuit, where it was noted that this process could be used for sensitive amplification \cite{naam08chirp}. Unlike the amplitude driven case, autoresonant excitation is by construction a non-equilibrium process, and in this letter we analyze the finite width of the threshold and the role of quantum fluctuations in determining the ultimate sensitivity of a measurement device which utilizes this effect.

The oscillator circuit consists of a tunnel junction embedded in the center of a linear cavity (see Fig.\ 1d), realized using a capacitively isolated section of coplanar waveguide transmission line that forms a 6 GHz Fabry-Perot resonator. This architecture allows us to engineer the electrical impedance shunting the junction and hence tune the oscillator quality factor ($Q$) and provides a simple means to couple microwave frequency signals into and out of the oscillator.  The equation of motion for the embedded system at large excitation differs from that of a physical pendulum; however, the autoresonant dynamics are unchanged.  We excite the circuit with a frequency chirp, created using a voltage controlled oscillator modulated by an arbitrary waveform generator. This signal is sent via heavily filtered coaxial lines to the chip, which is cooled to temperatures as low as 15 mK in a dilution refrigerator. The transmitted voltage is amplified at 2.3 K and demodulated using quadrature mixers at room temperature to obtain the magnitude relative to the drive.

We first demonstrate threshold behavior when the oscillator is driven with a phase continuous voltage pulse that starts at 6.075 GHz and decreases linearly in time to 5.775 GHz. The linear resonance frequency of the oscillator is  $\omega/(2\pi)= 5.987$ GHz with Q = 8200. Thus, both the start and stop frequencies are many linewidths away from resonance. The transmitted voltage ($\bar{V}$) developed in response to the drive is plotted in Figure 2a as a function of drive frequency and amplitude. The data are acquired by averaging 5,000 frequency sweeps at a fixed amplitude and then stepping its value over a 10 nV interval. As the drive becomes commensurate with the resonance frequency, oscillation amplitude begins to build. When the drive amplitude is weak, we observe that after a slight excitation the oscillator relaxes back to very small oscillations. On the other hand, for stronger drive, the oscillator builds up energy, behavior that is indicative of autoresonant phase-locking of the oscillations to the drive. 

    \begin{figure*}
\begin{center}
\includegraphics[angle = 0, width =.8 \textwidth]{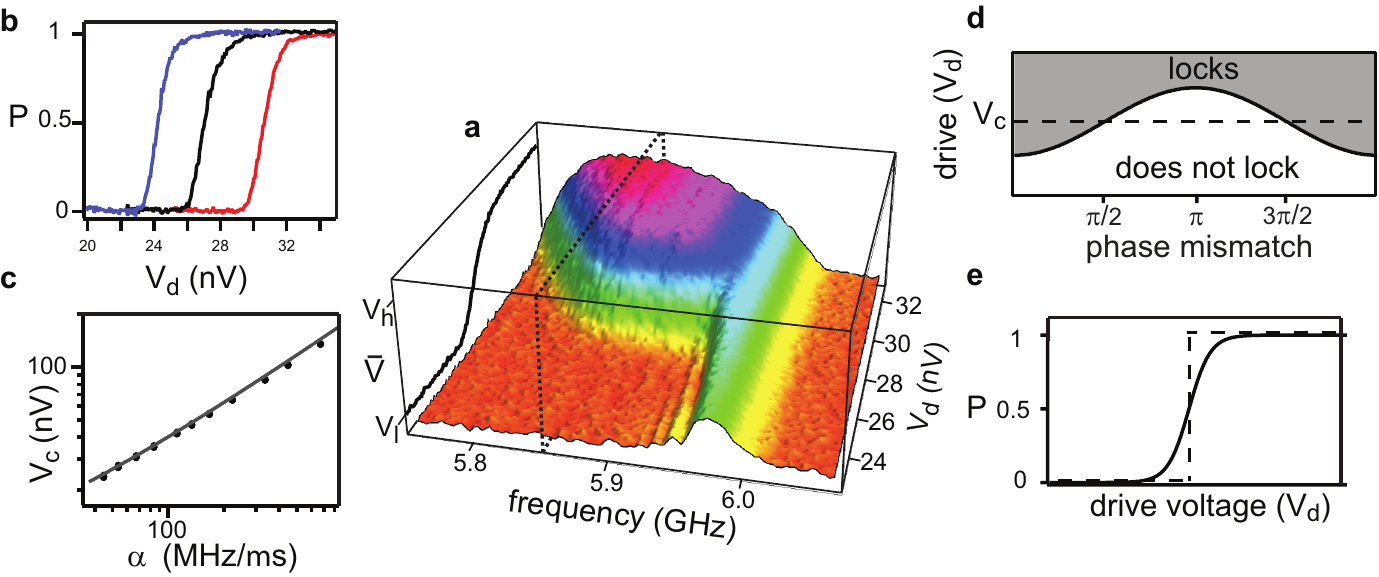}
\end{center}
\caption[mt chirp]{{\bf Autoresonance in the chirped pendulum.} {\bf a}, The average transmitted voltage $(\bar{V})$ versus drive frequency and drive voltage.   The drive was chirped at a rate of $50.6$ MHz/$\mu$s. The dashed line indicates the fixed frequency used for the slice (left panel), which shows the threshold for autoresonance to occur at $\vc \simeq 27$ nV. {\bf b}, The locking probability ascertained near the autoresonant threshold for chirp rates of 42.1 (blue), 50.6(black), and 63.2 (red) MHz/$\mu$s. {\bf c},  The threshold location, $\vc$ varies with the chirp rate and agrees with theoretical predictions \cite{naam08chirp}. {\bf d},  Cartoon of the locking/non-locking regions for finite amplitude (solid line) and zero (dashed line) fluctuation versus the phase mismatch between the drive and the oscillator. {\bf e},  Averaging over the initial amplitudes and phases of fluctuations broadens the transition  (solid line) from the zero fluctuation (dashed line) prediction.}
\end{figure*}

A constant frequency slice of the data in Figure 2a, shown in the left panel, shows the threshold for autoresonance at $\vc\simeq 27$ nV.  Here, the average junction voltage varies between a low level ($V_l$) corresponding to the unlocked events and a high level ($V_h$) corresponding to locked events.  Within the threshold, in a given frequency sweep the oscillator either locks into autoresonance or does not.  Because we average many sweeps, we extract the locking probability from this measurement, $P = (\bar{V}-V_l)/(V_h-V_l)$.   The locking probability versus drive power is shown for three different chirp rates in Figure 2b.  The threshold location, $\vc$, where $P=1/2$, is plotted versus chirp rate in Figure 2c, and agrees with theory \cite{naam08chirp}.

We now turn to the width of the autoresonant threshold. In the absence of fluctuations, the capture probability would be an infinitely sharp step at $\vc$ as indicated in Figures 2d and 2e. At the start of a chirp sequence, the resonator has an initial excitation that is the result of interactions with the fluctuating bath up to that time.  The threshold for a given initial amplitude $A_{0}$ is thus either augmented or diminished depending on initial the phase mismatch, $\Delta\phi$, between the drive signal and the oscillator motion, $\tilde{\vc} = \vc -\kappa  A_0\cos \Delta\phi$, where $\kappa=0.245$ is a prefactor determined by numerical simulations.  A finite threshold width is obtained by integrating the locking probability (see Fig.\ 2d) over all values of $\Delta \phi$ and over a thermal distribution of initial amplitudes at temperature $\teff$, where $k_B \teff =\frac{\hbar \omega}{2} \frac{1+ \exp(-\hbar \omega/k_B T)}{1-\exp(-\hbar \omega/k_B T)}$. The threshold width, defined as the inverse slope of the locking probability at $P=1/2$ is then $\dvc = 2 \kappa \sqrt{2\pi L \alpha k_B \teff}$, where $L$ is the total inductance of the resonator and $k_B$ is Boltzmann's constant \cite{bart09}.

 \begin{figure}
\begin{center}
\includegraphics[angle = 0, width = .45\textwidth]{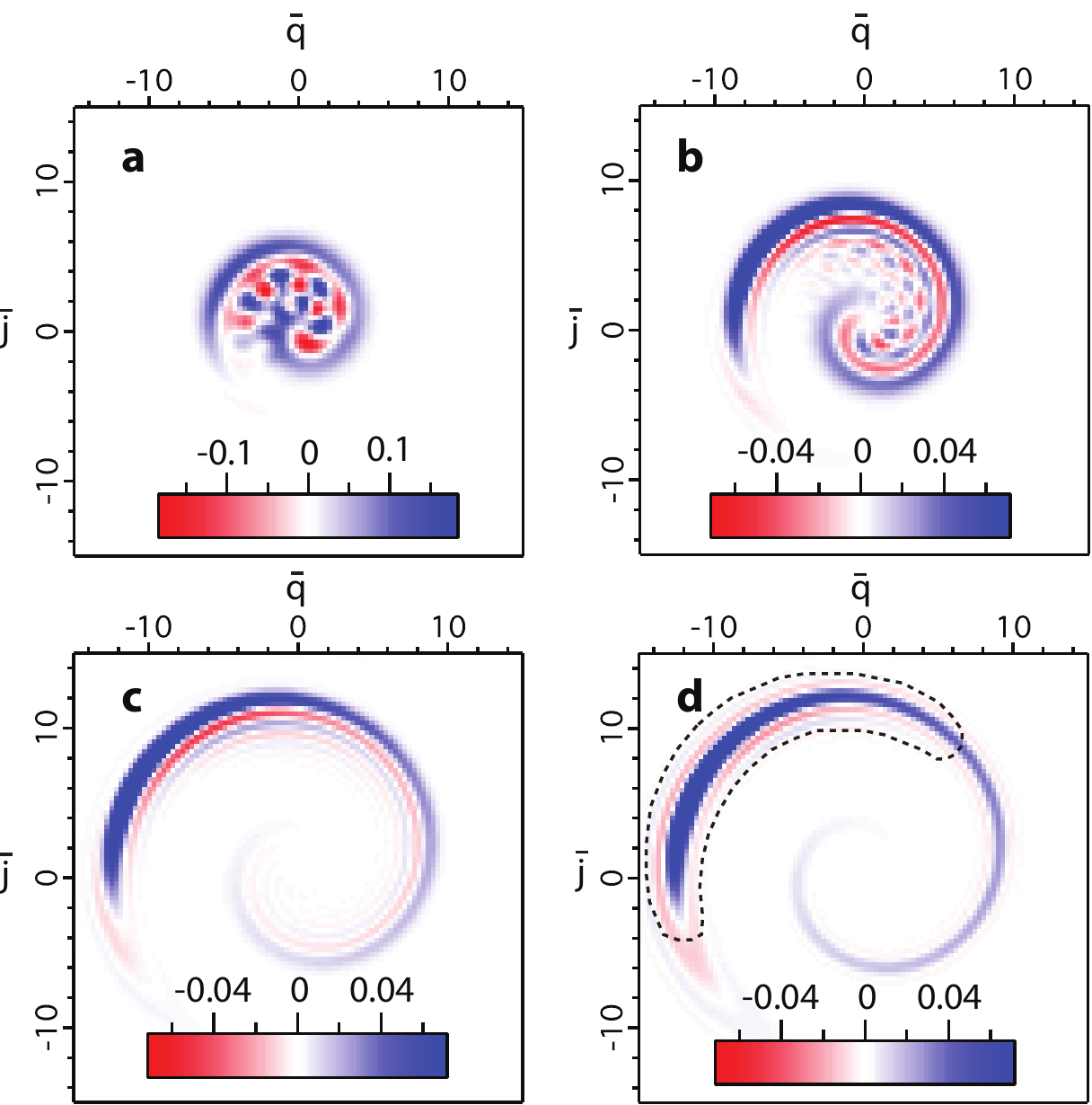}
\end{center}
\caption{{\bf Calculated Wigner distributions.}   Calculated Wigner distributions ($f(q,j,\tau=4215)$) for the quantum ({\bf a}-{\bf c}) and classical nonlinear oscillator ({\bf d}) as a function of the dimensionless charge $q$ and current $j$. {\bf a}-{\bf c} correspond to nonlinear energy level shifts that are respectively 100, 49, and 25 times larger than in the experiment.  The classical result in {\bf d} shows that approximately $90\ \%$ of the distribution is in the phase-locked state, as indicated by the dashed line. The tail outside this region relaxes to low amplitude if dissipation is included.  Negative values in the classical calculation result from the finite accuracy of the simulation.} \label{fig:wigner}
\end{figure}

To confirm the validity of this description when $k_B T\ll \hbar \omega$, we compute the dynamics in the presence of quantum fluctuations. The state of an oscillator can be described by its position and momentum, which for our electrical circuit correspond to the dimensionless charge ($q$) and current ($j$), respectively. In Figure 3, we plot the Wigner quasi-probability distribution that results after the ground state\textemdash a Gaussian centered about zero\textemdash is evolved using a voltage excitation 20 $\%$ greater than $\vc$. When the anharmonicity is large compared to dissipation, high excitation is achieved by climbing the ladder of accessible states, as opposed to a continuous classical evolution \cite{marc04ladder}. Panels a-c correspond to decreasing anharmonicity with energy level shifts 100, 49, and 25 times larger than our experimental parameters; panel d is the classical result with the majority of the population in the phase locked state, as indicated by the dashed line. The data in Figure 3c show nearly the same distribution as that of the classical calculation. For higher anharmonicity, the phase locked state remains qualitatively unchanged, but the population in the low amplitude state exhibits a characteristic quantum interference pattern indicating the participation of a relatively small number of discrete levels. Our sample parameters, however, correspond to an oscillator with 25 times \emph{weaker} anharmonicity than Figure 3c, and we thus expect the classical model for the transition width to be valid at all temperatures.

 \begin{figure}
\begin{center}
\includegraphics[angle = 0, width = .40\textwidth]{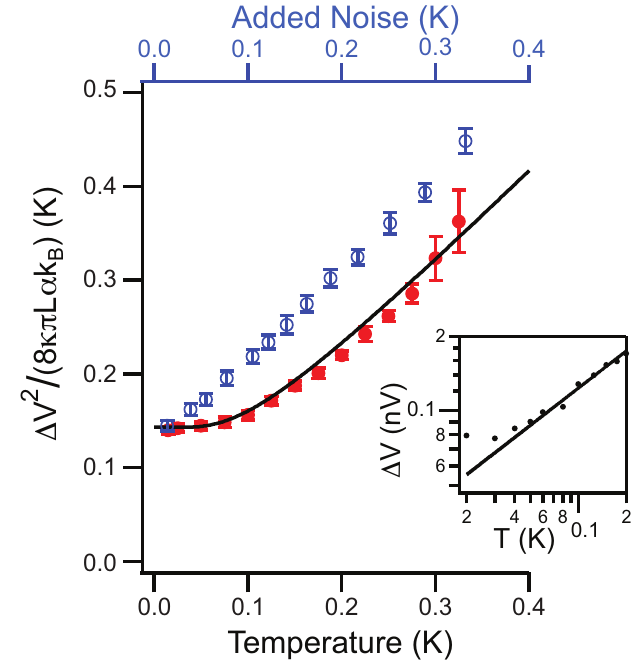}
\end{center}
\caption[width figure]{{\bf The autoresonance threshold width.}    The square of the threshold width, scaled to temperature units as $\Delta V^2/(8 \kappa \pi L \alpha k_B)$, and averaged across chirp rates ranging from 50 to 500 MHz/$\mu$s versus system temperature (red circles), saturates to the value predicted for quantum fluctuations.  The data are in agreement with our predictions, represented as an effective temperature, $\teff$, and shown as a solid black line.  White noise injected into the system causes the threshold width to increase $\propto T^{1/2}$ (blue open circles) with an offset given by the zero point quantum fluctuations in the resonator.  Inset: the threshold width for a junction embedded resonator at 1.6 GHz exhibits $\propto T^{1/2}$ scaling down to $T = 40$ mK. } \label{fig:kerrfigure}
\end{figure}

By varying the physical temperature of the dilution refrigerator we measured the transition width as a function of temperature.  To compare to theory we plot the square of the transition width, scaled to temperature units as $\Delta V^2/(8 \kappa \pi L \alpha k_B)$, versus temperature in Figure 4. For temperatures above 200 mK, we observe a clear $T^{1/2}$ dependence of the width. At the lowest temperatures, saturation is observed corresponding to $\teff = 144$ mK, which is precisely $\hbar \omega/(2k_B)$. The solid line indicates our theoretical prediction for the width with no adjustable parameters. To verify the quantum origin of this saturation, we plot the threshold width data for a 1.6 GHz resonator (Figure 4 inset). Here, we observe a $T^{1/2}$ scaling over the entire temperature range, with a suggestive flattening of the transition width at the lowest temperatures, where $k_B T \simeq \hbar \omega$.  

Keeping the sample at $T = 15$ mK, we also  injected white noise into the resonator to simulate a thermal bath. The results of this experiment are shown as the blue open circles in Figure 4.  The observed threshold width scales as $T^{1/2}$ and has an offset corresponding to the half quantum of zero point motion of the oscillator, corroborating again a quantum noise broadened threshold at the lowest temperature. Thus, quantum mechanics in our system enters only through the initial fluctuations, characterized by $\teff$, and the system otherwise can be treated classically \cite{Yurkbook}. 

Our experiment is an example of the interaction of an oscillator out of equilibrium  with a noise bath. The simple temperature scaling of the threshold width can potentially be harnessed for robust noise thermometry. Furthermore, the quantum saturation of this width,  which we note can be tuned by adjusting $\alpha$, ultimately sets the resolution of a digital detector based on autoresonance. Such a detector can be used for the readout of quantum bits.  In contrast to abrupt switching between metastable states \cite{naka09jba}, autoresonance involves continuous evolution to either the locked or unlocked state, raising interesting questions about the measurement backaction.

{\bf Methods}
The oscillator was fabricated from Al deposited on a high-resistivity Si wafer.  At low excitation power, a total quality factor of $Q = 8230$ resulted from internal losses characterized by $Q_{int} = 17200$ and coupling to the 50 $\Omega$ environment given by $Q_{ext} = 15800$.
 The critical power  for bifurcation was measured to be $P_c = -123$ dBm, corresponding to a junction critical current $I_0 = 1.8\ \mu$A and a total inductance $L=2.3$ nH.    At drive powers near $P_c$ we observed a decrease in the total quality factor of the oscillator to $Q = 6480$.

 The chirped pulse was created by driving a VCO  operating near 8 GHz with an arbitrary waveform generator and mixing it with a tone at 2.312 GHz. 
 The transmitted power from the sample was 
 mixed with the output of the VCO to form a fixed frequency signal at 2.312 GHz.  This signal was demodulated with a local oscillator at 2.412 GHz, and the I and Q quadratures were digitized at 500 MS/s.

 We modeled our oscillator via a weakly nonlinear Hamiltonian,
$
H = j^2/2 + q^2/2 -\beta q^4/4 + \varepsilon q \cos \phi_d,
$
on the phase space of dimensionless charge, $q$, normalized to $q_0= j_0/\omega$ and current, $j$, normalized to $j_0 = \sqrt{k_B \teff/L}$. The anharmonicity is given by $\beta = (\Phi_0 \omega^2 Q_0^2)/(6 L I_0^3)$, where $\Phi_0 = \hbar/2e$ is the flux quantum.  The dimensionless drive voltage $\varepsilon = V_d/(Lq_0 \omega^2)$ and the phase of the drive $\phi_d = \omega t - \alpha t^2/2$.  We numerically solved the dimensionless quantum Liouville equation for the Wigner function, $f(q, j, \tau)$, where $\tau = \omega t$:
 \begin{eqnarray}
\frac{\partial f}{\partial \tau} + j\frac{\partial f}{\partial q}-[(q-\beta q^3) + \varepsilon \cos \phi_d] \frac{\partial f}{\partial j} = \frac{\gamma^2 \beta q}{4 } \frac{\partial^3 f}{\partial j^3}.  \label{eq:l2}
\end{eqnarray}
 The parameter $\gamma = \hbar \omega/(k_B \teff)$ allows us to solve both the classical $\gamma\rightarrow 0$ and quantum $\gamma\rightarrow 2$ limits of the system.   We solved the Liouville equation using a standard pseudospectral method  \cite{canu54book} for the parameters in our experiment, but for  $\beta = n^2 \times 3.55 \times 10^{-6}$, $\varepsilon = 0.0246/n$,  and $\alpha = 10^{-6} \omega^2$ Hz$^2$, where $n=\{10, 7, 5\}$ correspond to nonlinear energy level shifts that are 100, 49 and 25 times that of the experiment, and drive strengths that are 20\% above the autoresonant threshold.  The classical evolution was approximated by solving the Liouville equation with $\gamma = 10^{-4}$.
 
 \vspace{.2in}
 % {\bf Acknowledgements}
We thank A. G. Shagalov, who developed the pseudospectral code used for solving the Liouville equation.
 This research was funded by the Office of the Director of National Intelligence (ODNI), Intelligence Advanced Research Projects Activity (IARPA), through the Army Research Office.  All statements of fact, opinion or conclusions contained herein are those of the authors and should not be construed as representing the official views or policies of IARPA, the ODNI, or the U.S. Government.  R.V. acknowledges funding from AFOSR under Grant No. FA9550-08-1-0104; I.B. and L.F. acknowledge support from the Israel Science Foundation under Grant No. 451/10.

%\bibliographystyle{naturemag}
%\bibliography{allrefs_2010_kwm}

\end{document}